\documentclass[letterpaper,twocolumn,10pt]{article}

\usepackage{usenix2019_v3}
\usepackage{filecontents}
\usepackage{amsmath}
\usepackage{mathtools}
\usepackage{cite}
\usepackage[english]{babel}
\usepackage{amssymb}
\usepackage{pifont}
\usepackage{textcomp}
\usepackage{tikz}
\usepackage{graphicx}
\usepackage{arydshln}
\usepackage{mathrsfs}
\usepackage{wasysym}
\usepackage{algorithm}
\usepackage{algorithmicx}
\usepackage{algpseudocode}
\usepackage{comment}
\usepackage[utf8]{inputenc}
\usepackage{multirow}
\usepackage{cleveref}
\usepackage{enumitem}
\usepackage{setspace}
\usepackage{url}
\usepackage{xspace}
\usepackage{accents}
\usepackage{ragged2e}
\usepackage{caption}
\usepackage{subcaption}
\usepackage{nth}
\usepackage{tightheaders}
\usepackage{blindtext}
\usepackage{booktabs} 
\usepackage{outlines}
\usepackage{color}

\newcommand{\fig}[1]{Figure~\ref{fig:#1}}
\newcommand{\tabl}[1]{Table~\ref{table:#1}}

\newcommand{\sect}[1]{\S~\ref{section:#1}}

\newcommand{\red}{}
\newcommand{\blue}{}

\begin{document}

\date{}

\title{\Large \bf Fault Tolerance for Service Function Chains}

\author{
{\rm Milad Ghaznavi}\\
University of Waterloo
\and
{\rm Elaheh Jalalpour}\\
University of Waterloo
\and
{\rm Bernard Wong}\\
University of Waterloo
\and
{\rm Raouf Boutaba}\\
University of Waterloo
\and
{\rm Ali Jos\'e Mashtizadeh}\\
University of Waterloo
} 

\maketitle

\begin{abstract}
Enterprise network traffic typically traverses a sequence of middleboxes forming a service function chain, or simply a chain.
Tolerating failures when they occur along chains is imperative to the availability and reliability of enterprise applications.
Making a chain fault-tolerant is challenging since, in the event of failures, the state of faulty middleboxes must be correctly and quickly recovered while providing high throughput and low latency.

In this paper, we introduce FTC, a system design and protocol for fault-tolerant service function chaining. FTC provides strong consistency with up to $f$ middlebox failures for chains of length $f+1$ or longer without requiring dedicated replica nodes. In FTC, state updates caused by packet processing at a middlebox are collected, piggybacked onto the packet, and sent along the chain to be replicated.
Our evaluation shows that compared with the state of art~\cite{rollback}, FTC improves throughput by 2--3.5$\times$ for a chain of two to five middleboxes.
\end{abstract}

\section{Introduction}
\label{section:introduction}

Middleboxes are widely deployed in enterprise networks, with each providing a specific dataplane function. These functions can be composed to meet high-level service requirements 
by passing traffic through an ordered sequence of middleboxes, forming a service function chain~\cite{rfc-sfc,simplefying}.
For instance, data center traffic commonly passes through an intrusion detection system, a firewall, and a network address translator before reaching the Internet~\cite{nfv-usecases}.

Providing fault tolerance for middleboxes is critical as their failures
have led to large network outages, significant financial losses, and left networks vulnerable to attacks~\cite{Potharaju2013, amazon-outage,netflix-outage,google-outage}. 
Existing middlebox frameworks~\cite{rollback,pico-replication,stateless,reinforce,chc} have focused on providing fault tolerance for individual middleboxes.
For a chain, they consider individual middleboxes as fault tolerant units that together form a fault tolerant chain.
This design introduces redundancies and overheads that can limit a chain's performance.

Independently replicating the state of each middlebox in a chain requires a large number replica servers, which can increase cost.
Part of that cost can be mitigated by having middleboxes share the same replica servers, although oversharing can affect performance.
More importantly, replication causes packets to experience more than twice its normal delay, since each middlebox synchronously replicates state updates before releasing a packet to the next middlebox~\cite{chc,stateless,pico-replication,reinforce}.

Current state-of-the-art middlebox frameworks also stall as they capture a consistent snapshot of their state leading to lower throughput and higher latency~\cite{rollback,pico-replication,reinforce}.
These stalls significantly increase latency with packets experiencing latencies from 400~$\mu s$ to 9~ms per middlebox compared to 10--100~$\mu s$ without fault tolerance~\cite{reinforce,pico-replication}.
When these frameworks are used in a chain, the stalls cause processing delays across the entire chain, similar to a {\em pipeline stall} in a processor.
As a result, we observed a $\sim$40\% drop in throughput for a chain of five middleboxes as compared to a single middlebox (see \sect{ftsfc-in-normal-operation}).

In this paper, we introduce a system called \emph{fault tolerant chaining} (FTC) that provides fault tolerance to an entire chain.
FTC
is inspired by {\em chain replication}~\cite{chain-replication} to efficiently provide fault tolerance.
At each middlebox, FTC collects state updates due to packet processing and piggybacks them onto the packet.
As the packet passes through the chain, FTC replicates piggybacked state updates in servers hosting middleboxes.
This allows each server hosting a middlebox to act as a replica for its predecessor middleboxes.
If a middlebox fails, FTC can recover the lost state from its successor servers.
For middleboxes at the end of the chain, FTC transfers and replicates their state updates in servers hosting middleboxes at the beginning of the chain. FTC does not need any dedicated replica servers.

We extend chain replication~\cite{chain-replication} to address challenges unique to a service function chain.
Unlike the original protocol where all nodes run an identical process,
FTC must support a chain comprised of different middleboxes processing traffic in the service function chain order.
Accordingly,
FTC allows all servers to process traffic
and replicate state.
Moreover, FTC's failure recovery instantiates a new middlebox at the failure position to maintain the service function chain order, rather than the traditional protocol that appends a new node at the end of a chain.

Furthermore, FTC improves the usability and performance of multicore middleboxes.
We introduce {\em packet transactions} to provide a simple programming model to develop multithreaded middleboxes that can
effectively make use of multiple cores.
Concurrent state updates to middlebox state result in non-deterministic behavior that is hard to restore.
A transactional model for state updates allows serializing concurrent state accesses that simplifies reasoning about both middlebox and FTC correctness.
The state of the art~\cite{rollback} relies on complex static analysis that supports unmodified applications, but can have worse performance when its analysis falls short.

FTC also tracks dependencies among transactions using \emph{data dependency vectors} that define a partial ordering of transactions. The partial ordering allows a replica to concurrently apply state updates from non-depend\-ent transactions to improve replication performance.
This approach has two major benefits compared to {\em thread-based} approaches
that allow concurrent state replication by replaying the operations of threads~\cite{rollback}.
First, FTC can support {\em vertical scaling} by replacing a running middlebox with a new instance with more CPU cores or failing over to a server with fewer CPU cores when resources are scarce during a major outage.
Second, it enables a middlebox and its replicas to run with a different number of threads.

FTC is implemented on Click~\cite{click} and uses the ONOS SDN controller~\cite{onos}.
We compare its performance with the state of the art in~\cite{rollback}.
Our results for a chain of two to five middleboxes show that FTC improves the throughput of the state of art~\cite{rollback} by 2$\times$ to 3.5$\times$ with lower latency per middlebox.

\section{Background}
\label{section:background}
A service function chain is an ordered sequence of middleboxes.
Today, following the {\em network function virtualization} (NFV) vision~\cite{nfv-whitepaper}, most middleboxes are implemented as software running on commodity hardware.

\begin{figure}[t]
    \centering
    \includegraphics[width=.4\textwidth]{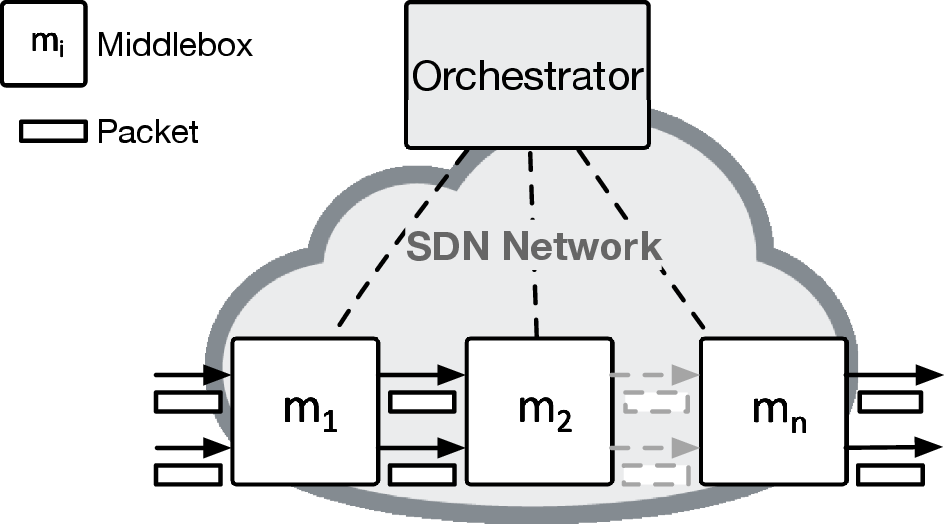}
    \caption{Service function chain model in NFV}
    \label{fig:sfc-model}
\end{figure}

In an NFV environment, as shown in \fig{sfc-model}, an orchestrator manages and steers traffic through a chain of middleboxes.
Each middlebox runs multiple threads and is equipped with a multi-queue network interface card (NIC)~\cite{routerbricks,comb,locality-multi-core}.
A thread receives packets from a NIC's input queue and sends packets to a NIC's output queue.
\fig{sfc-model} shows two threaded middleboxes processing two traffic flows.

Stateful middleboxes keep dynamic state for packets that they process~\cite{flow-associated, hilti}.
For instance, a stateful firewall filters packets based on statistics that it collects for network flows~\cite{conn-netfliter},
and a network address translator (NAT) maps internal and external addresses using a flow table~\cite{rfc3022,rfc2993}.

Middlebox state can be \emph{partitionable} or \emph{shared}~\cite{rfc2979,conn-netfliter,OpenNF,pico-replication}.
Partitionable state variables describe the state of a single traffic flow (e.g., MTU size and timeouts in stateful firewalls~\cite{conn-netfliter,rfc2979}) and are only accessed by a single middlebox thread.
Shared state variables are for a collection of flows, and multiple middlebox threads query and update them (e.g., port-counts in an intrusion detection system).

A stateful middlebox is subject to both hardware and software failures that can cause the loss of its state~\cite{Potharaju2013,rollback}.
We model these failures as {\em fail-stop} in which failures are detectable, and failed components are not restored.

\subsection{Challenges}
\label{section:challenges}
To recover from a middlebox failure, traffic must be rerouted to a redundant middlebox where the state of the failed middlebox is restored.
State replication has two challenges that affect middlebox performance.

First, in a multithreaded middlebox, the order in which interleaving threads access shared state is non-deterministic.
Parallel updates can lead to observable states that are hard-to-restore.
{\blue The difficulty in achieving high performance multithreaded middleboxes is how we capture this state for recovery.}
One approach to accommodate non-determinism is to log any state read and write, which allows restoring any observable state from the logs~\cite{rollback}.
However, this complicates the failure recovery procedure because of record/replay, and leads to high performance overheads during normal operation.

Second, to tolerate $f$ failures, a packet is released \emph{only} when at least $f+1$ replicas acknowledge that state updates due to processing of this packet are replicated.
In addition to increasing latency, synchronous replication reduces throughput since expensive coordinations between packet processing and state replication are required for consistency (e.g., pausing packet processing until replication is acknowledged~\cite{pico-replication,reinforce,stateless,chc}).
The overhead of this synchrony for a middlebox depends on where its replicas are located, and how state updates are transferred to these locations.
For a solution designed for individual middleboxes,
the overheads can accumulate for each middlebox of a chain.

\subsection{Limitations of Existing Approaches}
\label{section:fault-tolerance-for-middleboxes}
Existing middlebox frameworks provide fault tolerance for individual middleboxes.
Using these frameworks for a chain whose middleboxes are deployed over multiple servers significantly impacts the chain's performance.
These frameworks use one of two approaches.

A class of frameworks take snapshots of middlebox state for state replication~\cite{pico-replication,reinforce,rollback}.
While taking snapshot, middlebox operations are stalled for consistency.
These frameworks take snapshots at different rates.
They take snapshots per packet or packet-batch introducing 400~$\mu$s to 8--9~ms of per packet latency overhead~\cite{pico-replication,reinforce}.
Periodic snapshots (e.g., at every 20--200~ms intervals) can cause periodic latency spikes up to 6~ms~\cite{rollback}.
We measure that per middlebox snapshots cause 40\% throughput drop going from a single middlebox to a chain of five middleboxes (see \sect{ftsfc-in-normal-operation}).

Other frameworks~\cite{stateless,chc} redesign
middleboxes to separate and push state into a fault tolerant backend data store.
This separation incurs high performance penalties.
Accessing state takes at least a round trip delay.
Moreover, a middlebox can release a packet only when it receives an acknowledgement from the data store that relevant state updates are replicated.
Due to such overheads,
the middlebox throughput
can drop by $\sim$60\%~\cite{stateless} and reduce to
0.5~Gbps (for packets with 1434~B median size)~\cite{chc}

\section{System Design Overview}
\label{section:fault tolerant-service-function-chaining}

The limitations of existing work lead us to design fault tolerant chaining (FTC); a new approach that replicates state along the chain to provide fault tolerance.

\subsection{Requirements}
\label{section:requirements}

We design FTC to provide fault tolerance for a wide variety of middleboxes.
FTC adheres to four requirements:

\paragraph{Correct recovery:}
FTC ensures that the middlebox behavior after a failure recovery is consistent with the behavior prior to the failure~\cite{optimistic-recovery}.
To tolerate $f$ failures, a packet can only be released outside of a chain once all necessary information needed to reconstruct the internal state of all middleboxes is replicated to $f+1$ servers.

\paragraph{Low overhead and fast failure recovery:}
Fault tolerance for a chain must come with low overhead.
A chain processes a high traffic volume and middlebox state can be modified very frequently.
At each middlebox of a chain, latency should be within 10 to 100~$\mu$s~\cite{rollback}, and the fault tolerance mechanism must support accessing variables 100~k to 1~M times per second~\cite{rollback}.
Recovery time must be short enough to prevent application outages.
For instance, highly available services timeout in just a few seconds~\cite{ms-failover-tuning}.

\paragraph{Resource efficiency:}
Finally, the fault tolerance solution should be resource efficient.
To isolate the effect of possible failures, replicas of a middlebox must be deployed on separate physical servers.
We are interested in a system that dedicates the fewest servers to achieve a fixed replication factor.

\subsection{Design Choices}

We model packet processing as a transaction.
FTC carefully collects updated values of state variables modified during a packet transaction and appends them to the packet.
As the packet passes through the chain, FTC replicates piggybacked state updates in servers hosting the middleboxes.

\paragraph{Transactional packet processing:}
To accommodate non-determinism due to concurrency, we model the processing of a packet as a transaction, where concurrent accesses to shared state are serialized to ensure that consistent state is captured and replicated.
{\blue
In other systems, the interleaved order of lock acquisitions and state variable updates between threads is non-deterministic, yet externally observable.
Capturing and replaying this order is complex and incurs high performance overheads~\cite{rollback}.
FTC uses transactional packet processing to avoid the complexity and overhead.
}

This model is easily adaptable to hybrid transactional memory, where we can take advantage of the hardware support for transactions~\cite{DAVEDICE}.
This allows FTC to use modern hardware transactional memory for better performance, when the hardware is present.

We also observe that this model does not reduce concurrency in popular middleboxes.
First, these middleboxes already serialize access to state variables for correctness.
For instance, a load balancer and a NAT ensure {\em connection persistence} (i.e., a connection is always directed to a unique destination) while accessing a shared flow table~\cite{rfc3234,rfc3022}.
Concurrent threads in these middleboxes must coordinate to provide this property.

Moreover, most middleboxes share only a few state variables~\cite{chc,stateless}.
Kablan et al. surveyed five middleboxes for their access patterns to state~\cite{stateless}.
These middleboxes mostly perform only one or two read/write operations per packet.
The behavior of these middleboxes allow packet transactions to run concurrently most of the time.

\paragraph{In-chain replication:}
Consensus-based state replication~\cite{lamport2001paxos,raft} requires $2f+1$ replicas for each middlebox to reliably detect and recover from $f$ failures.
A high-availability cluster approach requires $f+1$ replicas as it relies on a fault tolerant coordinator for failure detection.
For a chain of $n$ middleboxes, these schemes need $n\times(2f+1)$ and $n\times(f+1)$ replicas.
Replicas are placed on separate servers, and a {\em na{\"i}ve} placement requires the same number of servers.

FTC observes that packets already flow through a chain; each server hosting a middlebox of the chain can serve as a replica for the other middleboxes.
Instead of allocating dedicated replicas, FTC replicates the state of middleboxes across the chain.
In this way, FTC tolerates $f$ failures without the cost of dedicated replica servers.

\paragraph{State piggybacking:}
To replicate state modified by a packet, existing schemes send separate messages to replicas.
In FTC,
a packet carries its own state updates.
State piggybacking is possible, as a small number of state variables~\cite{Statealyzer} are modified with each packet.
Since state updated during processing a packet is replicated in servers hosting the chain, relevant state is already transferred and replicated when the packet leaves the chain.

\paragraph{No checkpointing and no replay:}
FTC replicates state values at the granularity of packet transactions, rather than taking snapshots of state or replaying packet processing operations.
During normal operation, FTC removes state updates that have been applied in all replicas to bound memory usage of replication.
Furthermore, replicating the values of state variables allows for fast state recovery during failover.

\paragraph{Centralized orchestration:}
In our system, a central orchestrator manages the network and chains.
The orchestrator deploys fault tolerant chains, reliably monitors them, detects their failures, and initiates failure recovery.
The orchestrator functionality is provided by a fault tolerant SDN controller~\cite{onos,ravana,scl}.
After deploying a chain, the orchestrator is not involved in normal chain operations to avoid becoming a performance bottleneck.

In the following sections, we first describe our protocol for a single middlebox in \sect{ftc-middlebox}, then we extend this protocol for a chain of middleboxes in \sect{ftc-chain}.

\section{FTC for a Single Middlebox}
\label{section:ftc-middlebox}
In this section, we present our protocol for a single middlebox.
We first describe our protocol with a single threaded middlebox where state is replicated by single threaded replicas.
We extend our protocol to support multithreaded middleboxes and multithreaded replication in \sect{packet-transactions-for-concurrent-packet-processing} and \sect{vector-clock}.

\subsection{Middlebox State Replication}
\label{section:chain-replication-for-fault-tolerance}
We adapt the {\em chain replication} protocol \cite{chain-replication} for middlebox state replication.
For reliable state transmission between servers, FTC uses \emph{sequence numbers}, similar to TCP, to handle out-of-order deliveries and packet drops within the network.

\begin{figure}[t]
    \centering
    \includegraphics[width=0.47\textwidth]{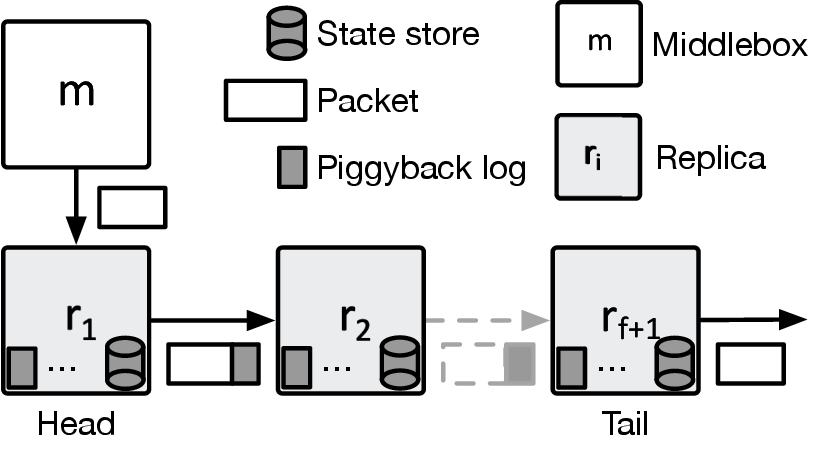}
    \caption{Normal operation for a single middlebox.
    {\normalfont The head and middlebox reside in the same server. The head tracks state updates due to middlebox packet processing and appends a piggyback log containing these updates to the packet. As the packet passes through the chain, other replicas replicate the piggyback log and apply the carried state updates to their state stores. Finally, the tail strips the piggyback log and releases the packet.}}
    \label{fig:middlebox-normal-operation}
\end{figure}

\fig{middlebox-normal-operation} shows our protocol for providing fault tolerance for a middlebox.
FTC replicates the middlebox state
in $f+1$ replicas during normal middlebox operations.
Replicas $r_1,\dots,r_{f+1}$ form the {\em replication group} for middlebox $m$ where $r_1$ and $r_{f+1}$ are called the {\em head} and {\em tail} replicas.
Each replica is placed on a separate server whose failure is isolated.
With state replicated in $f+1$ replicas, the state remains available even if $f$ replicas fail.

The head is co-located with the middlebox in the same server.
The middlebox state is separated from the middlebox logic and is stored in the head's {\em state store}.
The head provides a state management API for the middlebox to read and write state during packet processing.

\paragraph*{Normal operation of protocol:}
As shown in \fig{middlebox-normal-operation}, the middlebox processes a packet, and the head constructs and appends a {\em piggyback log} to the packet.
The piggyback log contains a sequence number and a list of state updates during packet processing.
As the packet traverses the chain, each subsequent replica replicates the piggyback log and applies the state updates to its state store.
After replication, the tail strips the piggyback log and releases the packet.

The head tracks middlebox updates to state using a monotonically increasing sequence number.
After a middlebox finishes processing a packet, the head increments its sequence number only if state was modified during packet processing.
The head appends the state updates (i.e.,~state variables modified in processing the packet and their updated values) and sequence number to the packet as a piggyback log.
If no state was updated, the head adds a no-op piggyback log.
The head then forwards the packet to the next replica.

Each replica continuously receives packets with piggyback logs.
If a packet is lost, a replica requests its predecessor to re-transmit the piggyback log with the lost sequence number.
A replica keeps the largest sequence number that it has received in order (i.e., the replica has already received all piggyback logs with preceding sequence numbers).
Once all prior piggyback logs are received, the replica applies the piggyback log to its local state store and forwards the packet to the next replica.

The tail replicates state updates, strips the piggyback log from the packet, and releases the packet to its destination.
Subsequently, the tail periodically disseminates its largest sequence number to the head.
The sequence number is propagated to all replicas so they can prune their piggyback logs up to this sequence number.

\paragraph*{Correctness:}
\label{section:mb-single-thread-correctness}
Each replica replicates the per-packet state updates in order.
As a result, when a replica forwards a packet, it has replicated all preceding piggyback logs.
Packets also pass through the replication group in order.
When a packet reaches a replica, prior replicas have replicated the state updates carried by this packet.
Thus, when the tail releases a packet,
the packet has already traversed the entire replication group.
The replication group has $f+1$ replicas allowing FTC to tolerate $f$ failures.

\paragraph*{Failure recovery:}
\label{section:recovery-mb}
{\blue FTC relies on a fault tolerant orchestrator to reliably detect failures.
Upon failure detection, the replication group is repaired in three steps:}
adding a new replica,
recovering the lost state from an alive replica,
and steering traffic through the new replica.

In the event of a head failure, the orchestrator instantiates a new middlebox instance and replica, as they reside on the same server.
The orchestrator also informs the new replica about other alive replicas.
If the new replica fails, the orchestrator restarts the recovery procedure.

Selecting a replica as the source for state recovery depends on how state updates propagate through the chain.
We can reason about this using the \emph{log propagation invariant}:
for each replica except the tail, its successor replica has the same or prior state, since piggyback logs propagate in order through the chain.

If the head fails, the new replica retrieves the state store, piggyback logs, and sequence number from the immediate successor to the head.
If other replicas fail, the new replica fetches the state from the immediate predecessor.

To ensure that the log propagation invariant holds during recovery,
the replica that is the source for state recovery discards any out-of-order packets that have not been applied to its state store and will no longer admit packets in flight.
If the contacted replica fails during recovery, the orchestrator detects this failure and re-initializes the new replica with the new set of alive replicas.

Finally, the orchestrator updates routing rules in the network to steer traffic through the new replica.
If multiple replicas have failed, the orchestrator waits until all new replicas acknowledge that they have successfully recovered the state.
Then, the orchestrator updates the necessary routing rules from the tail to the head.

\subsection{Concurrent Packet Processing}
\label{section:packet-transactions-for-concurrent-packet-processing}
To achieve higher performance, we augment our protocol to support multithreaded packet processing and state replication in the middlebox and the head.
Other replicas are still single threaded. Later in \sect{vector-clock}, we will support multithreaded replications in other replicas.

In concurrent packet processing, multiple packets are processed in interleaving threads.
The threads can access the same state variables in parallel.
To accommodate this parallelism,
FTC
must consistently track parallel state updates.
We introduce \emph{transactional packet processing} that effectively serializes packet processing.
This model supports concurrency if packet transactions access disjoint subsets of state.

\paragraph*{Transactional Packet Processing:}
\label{section:concurrent-packet-processing-and-state-logging}
{\red In concurrent packet processing, the effects on state variables must be  serializable.}
Further, state updates must be applied to replicas in the same order so that the system can be restored to a consistent state during failover.
To support this requirement, replay based replication systems, such as FTMB~\cite{rollback}, track all state accesses, including state reads, which can be challenging to perform efficiently.

In transactional packet processing, 
state reads and writes by a packet transaction have no impact on another concurrently processed packet.
This isolation allows us to only keep track of the relative order between transactions, without needing to track all state variable dependencies.

We realize this model by implementing a {\em software transactional memory} (STM) API for middleboxes.
When a packet arrives, the runtime starts a new packet transaction in which multiple reads and writes can be performed.
Our STM API uses {\em fine grained strict two phase locking} (similar to~\cite{bib:TL2}) to provide serializability.
Our API uses a {\em wound-wait scheme} that aborts transaction to prevent possible deadlocks if a lock ordering is not known in advance.
An aborted transaction is immediately re-executed.
The transaction completes when the middlebox releases the packet.

Using two phase locking, the head runtime acquires necessary locks during a packet transaction.
We simplify lock management using {\em state space partitioning}, by using the hash of state variable keys to map keys to partitions, each with its own lock.
The state partitioning is consistent across all replicas, and to reduce contention, the number of partitions is selected to exceed the maximum number of CPU cores.

At the end of a transaction, the head atomically increments its sequence number only if state was updated during this packet transaction.
Then, the head constructs a piggyback log containing the state updates and the sequence number.
After the transaction completes, the head appends the piggyback log to the packet and forwards the packet to the next replica.

\paragraph*{Correctness:}
Due to {\em mutual exclusion}, when a packet transaction includes an updated state variable in a piggyback log, {\em no} other concurrent transaction has modified this variable, thus the included value is consistent with the final value of the packet transaction.
The head's sequence number maps this transaction to a {\em valid} serial order.
Replicated values are consistent with the head, because replicas apply state updates of the transaction in the sequence number order.

\subsection{Concurrent State Replication}
\label{section:vector-clock}
Up to now FTC provides concurrent packet processing but does not support concurrent replication.
The head uses a single sequence number to determine a total order of transactions that modify state partitions.
This total ordering
eliminates multithreaded replication at successor replicas.

To address the possible replication bottleneck, we introduce {\em data dependency vectors} to support concurrent state replication.
Data dependency tracking is inspired by the {\em vector clocks} algorithm~\cite{vector-clock}, but rather than tracking points in time when events happen for processes or threads, FTC tracks the points in time when packet transactions modify state partitions.

This approach provides more flexibility compared to tracking dependencies between threads and replaying their operations to replicate the state~\cite{rollback}.
First, it easily supports vertical scaling as a running middlebox can be replaced with a new instance with different number of CPU cores.
Second, a middlebox and its replicas can also run with different number of threads.
The state-of-the-art~\cite{rollback} requires the same number of threads with a one-to-one mapping between a middlebox and its replicas.

\paragraph{Data dependency vectors:}
We use data dependency vectors to determine a partial order of transactions in the head.
Each element of this vector is a sequence number associated to a state partition.
A packet piggybacks this partial order to replicas enabling them to replicate transactions with more concurrency; a replica can apply and replicate a transaction in a different serial order that is still equivalent to the head.

\begin{figure}[t]
    \centering
    \includegraphics[width=0.475\textwidth]{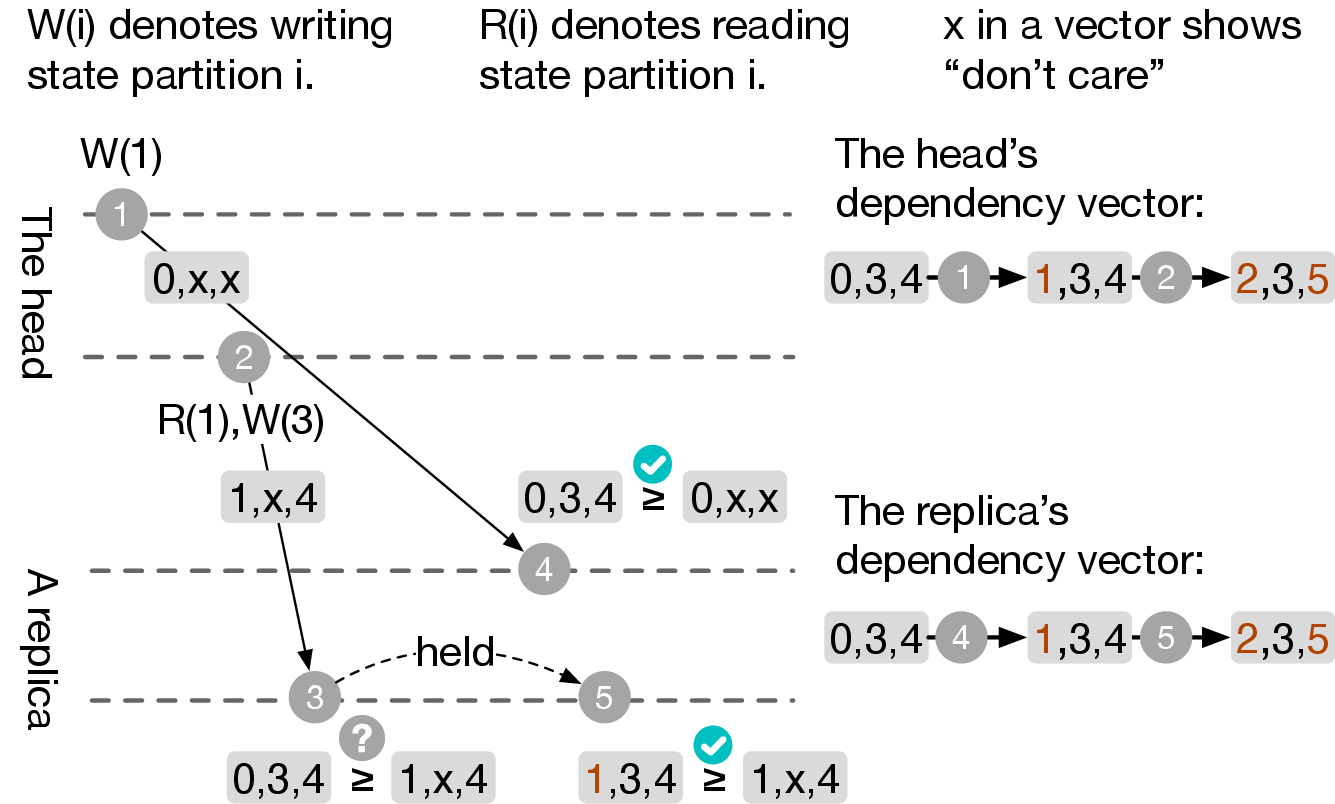}
    \caption{Data dependency vectors.
    {\normalfont The head and the replica run two threads and maintain a dependency vector for three state partitions.}}
    \label{fig:vc-at-head}
\end{figure}

The head keeps a data dependency vector and serializes parallel accesses
to this vector using the same state partition locks from our transactional packet processing.
The head maintains its dependency vector using the following rules.
A read-only transaction does {\em not} change the vector.
For other transactions, the head increments the sequence number of a state partition that received any read or write.

In a piggyback log, we replace the sequence number with a dependency vector that represents the effects of a transaction on state partitions.
If the transaction does not access a state partition, the head uses a ``{\em don't-care}'' value for this partition in the piggyback log.
The head obtains the sequence number of other partitions from the head's dependency vector before incrementing their sequence numbers.

Each successor replica keeps a dependency vector $MAX$ that tracks the latest piggyback log that it has replicated in order, i.e., it has already received all piggyback logs prior to $MAX$.
In case a packet is lost, a replica requests its predecessor to retransmit missing piggyback logs.

Upon receiving a packet, a replica compares the piggybacked dependency vector with its $MAX$.
The replica ignores state partitions with ``don't care'' from this comparison.
Once all prior piggyback logs have been received and applied, the replica applies and replicates the piggyback log.
For other state partitions, the replica increments their associated sequence numbers in $MAX$. 

\paragraph{Example:}
\fig{vc-at-head} shows an example of using data dependency vectors in the head and a successor replica with two threads.
The head and the replica begin with 
the same dependency vector
for three state partitions.
First, the head performs a packet transaction that writes to state partition 1 and increments the associated sequence number.
The piggyback log belonging to this transaction contains ``don't care'' value for state partitions 2 and 3 (denoted by {\tt x}), since the transaction did not read or write these partitions.
Second, the head performs another transaction and forwards the packet with a piggyback log.

Third, as shown the second packet arrives to the replica before the first packet.
Since the piggybacked dependency vector is out of order, the replica holds the packet.
Fourth, the the first packet arrives. Since the piggybacked vector is in order, the replica applies the piggyback log and updates its local dependency vector accordingly.
Fifth, by applying the piggyback log of the first packet, the replica now can apply the piggyback log of the held packet.

\section{FTC for a Chain}
\label{section:ftc-chain}

\begin{figure}[t]
    \centering
    \includegraphics[width=.475\textwidth]{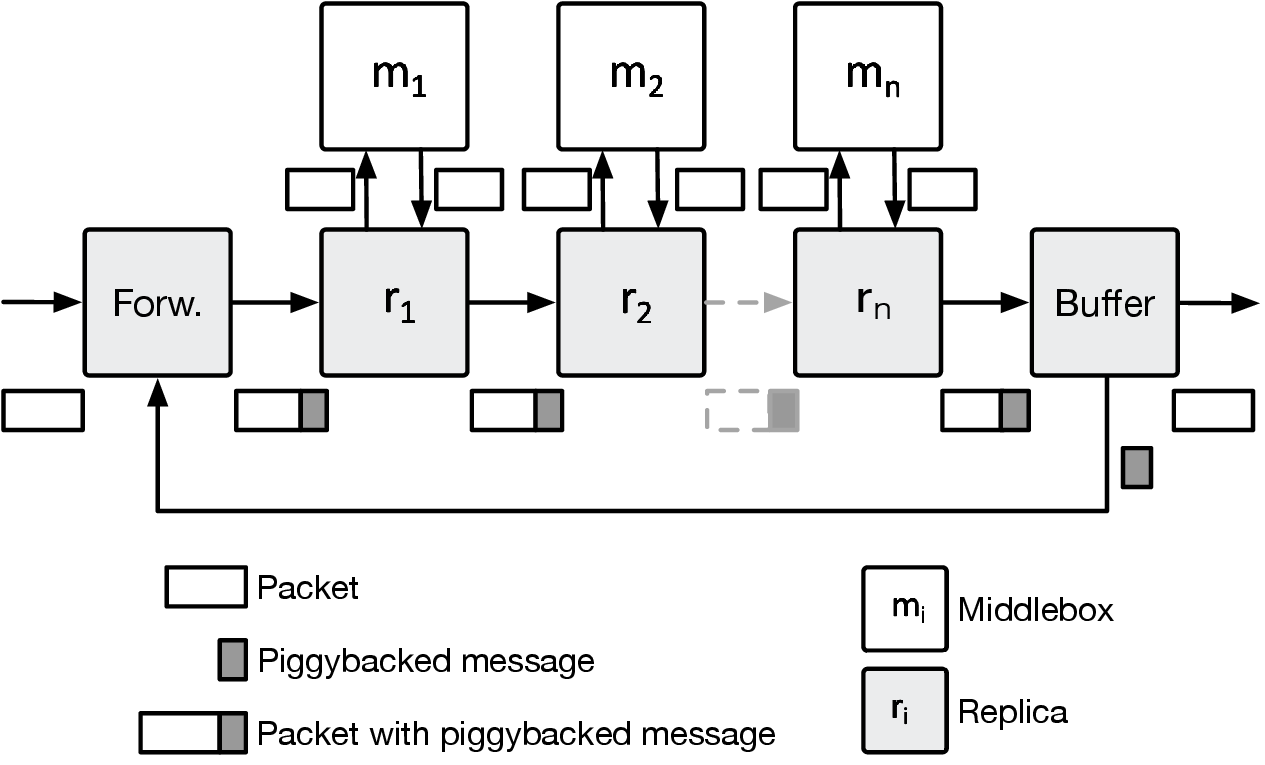}
    \caption{Normal operation for a chain. \normalfont{A middlebox $m_i$ and its head replica $r_i$ reside in the same server.
    The forwarder and buffer are located in the first and last servers. 
    A replica $r_i$ tracks state updates of a middlebox $m_i$ and adds a piggyback message including the state updates to the packet.
    As the packet traverses the chain, each replica replicates relevant carried state updates.
    Replicas at the beginning of the chain replicate for middleboxes at the end of the chain.
    The buffer withholds the packet from release until the state updates of middleboxes at the end of the chain are replicated.
    The buffer transfers the piggyback message to the forwarder that adds it to incoming packets for state replication.}}
    \label{fig:packet-stream}
\end{figure}
In this section, we describe our protocol for a chain to enable every middlebox to replicate the chain's state while processing packets.
To accomplish this, we extend the original chain replication protocol~\cite{chain-replication} during both normal operation and failure recovery.

A chain consists of different middlebox applications. Thus, FTC must allow different middleboxes to run across the chain, while the original chain replication protocol supports running an identical process across the nodes.
FTC's failure recovery instantiates a new middlebox at the failure position to maintain the chain order, while the traditional protocol appends a new node at the end of a chain.

\fig{packet-stream} shows our protocol for a chain of $n$ middleboxes.
Our protocol can be thought of as running $n$ instances (per middlebox) of the protocol developed earlier in \sect{ftc-middlebox}.
FTC places a replica per each middlebox.
Replicas form $n$ replication groups, each of which provides fault tolerance for a single middlebox.

Viewing a chain as a \emph{logical ring}, the replication group of a middlebox consists of a replica and its $f$ succeeding replicas.
Instead of being dedicated to a single middlebox, a replica is shared among $f+1$ middleboxes and maintains a state store for each of them.
Among these middleboxes, a replica is the head of one replication group and the tail of another replication group.
For instance in \fig{packet-stream}, if $f = 1$ then the replica $r_1$ is in the replication groups of middleboxes $m_1$ and $m_n$, and $r_2$ is in the replication groups of $m_1$ and $m_2$.
Subsequently, the replicas $r_n$ and $r_1$ are the head and the tail of middlebox $m_n$.

FTC adds two additional elements, the {\em forwarder} and {\em buffer} at the ingress and egress of a chain.
The forwarder and buffer are also multithreaded, and are collocated with the first and last middleboxes.
The buffer holds a packet until the state updates associated with all middleboxes of the chain have been replicated.
The buffer also forwards state updates to the forwarder for middleboxes with replicas at the beginning of the chain.
The forwarder adds state updates from the buffer to incoming packets before forwarding the packets to the first middlebox.

\subsection{Normal Operation of Protocol}
\label{section:chain-normal-operation}
\fig{packet-stream} shows the normal operation of our protocol.
The forwarder receives incoming packets from the outside world and {\em piggyback messages} from the buffer.
A piggyback message contains middlebox state updates.
As the packet passes through the chain, a replica detaches and replicates the relevant parts of the piggyback message and applies associated state updates to its state stores.
A replica $r_i$ tracks the state updates of a middlebox $m_i$ and updates the piggyback message to include these state updates.
Replicas at the beginning of the chain replicate for middleboxes at the end of the chain.
The buffer withholds the packet from release until the state updates of middleboxes at the end of the chain are replicated.
The buffer transfers the piggyback message to the forwarder that adds it to incoming packets for state replication.

The forwarder receives incoming packets from outside world and piggyback messages from the buffer.
A piggyback message consists of a list of piggyback logs and a list of {\em commit vectors}.
The tail of each replication group appends a commit vector to announce the latest state updates that have been replicated $f+1$ times for the corresponding middlebox.

Each replica constantly receives packets with piggyback messages.
A replica detaches and processes a piggyback message before the packet transaction.
As mentioned before, each replica is in the replication group of $f$ preceding middleboxes.
For each of them, the replica maintains a dependency vector $MAX$ to track the latest piggyback log that it has replicated in order.
The replica processes a relevant piggyback log from the piggyback message
as described in~\sect{vector-clock}.
Once all prior piggyback logs are applied, the replica replicates the piggyback log, applies state updates to the associated state store, and updates the associated dependency vector $MAX$.

Once the middlebox finishes the packet transaction,
the replica updates and reattaches the piggyback message to the packet, then forwards the packet.
For the replication group where the replica is the head,
it adds a piggyback log containing the state updates of processing the packet.
If the replica is a tail in the replication group of a middlebox $m$,
it removes the piggyback log belonging to middlebox $m$ to reduce the size of the piggyback message.
The reason is that a tail replicates the state updates of $m$ for $f+1$-th time.
Moreover, it attaches its dependency vector $MAX$ of middlebox $m$ as a commit vector.
Later by reading this commit vector,
the buffer can safely release held packets.
Successor replicas also use this commit vector to prune obsolete piggyback logs.

To correctly release a packet, the buffer requires that the state updates of this packet are replicated, specifically for each middlebox with a preceding tail in the chain.
The buffer withholds a packet from release until an upcoming packet piggybacks commit vectors that confirm meeting this requirement.
Upon receiving an upcoming packet, the buffer processes the piggybacked commit vectors
to release packets held in the memory.

Specifically, let $m$ be a middlebox with a preceding tail, and $V_2$ be the end of updated range from a piggyback log of a held packet belonging to $m$.
Once the commit vector of each $m$ from an upcoming packet shows that all state updates prior to and including $V_2$ have been replicated, the buffer releases the held packet and frees its memory.

\paragraph{Other considerations:}
There may be time periods that a chain receives no incoming packets. In such cases, the state is not propagated through the chain, and the buffer does not release packets.
To resolve this problem, the forwarder keeps a timer to receive incoming packets.
Upon the timeout, the forwarder sends a {\em propagating packet} carrying a piggyback message it has received from the buffer.
Replicas do not forward a propagating packet to middleboxes. 
They process and update the piggyback message as described before and forward the packet along the chain.
The buffer processes the piggyback message to release held packets.

Some middlebox in a chain can filter packets (e.g., a firewall may block certain traffic), and consequently the piggybacked state is not passed on.
For such a middlebox, its head generates a propagating packet to carry the piggyback message of a filtered packet.

Finally, if the chain length is less than $f+1$, we extend the chain by adding more replicas prior to the buffer.
These replicas only process and update piggyback messages.

\subsection{Failure Recovery}
Handling the failure of the forwarder or the buffer is straightforward.
They contain only soft state, and spawning a new forwarder or a new buffer restores the chain.

The failure of a middlebox and its head replica is not isolated, since they reside on the same server.
If a replica fails, FTC repairs $f+1$ replication groups as each replica replicates for $f+1$ middleboxes.
The recovery involves three steps: spawning a new replica and a new middlebox, recovering the lost state from other alive replicas, and steering traffic through the new replica.

After spawning a new replica,
the orchestrator informs
it
about the list of replication groups in which the failed replica was a member.
For each of these replication group, the new replica runs an independent state recovery procedure as follows.
If the failed replica was the head of a replication group,
the new replica retrieves the state store and the dependency vector $MAX$ from the immediate successor in this replication group.
The new replica restores the dependency matrix of the failed head by setting each of its row to the retrieved $MAX$.
For other replication groups, the new replica fetches the state from the immediate predecessors in these replication groups.

Once the state is recovered, the new replica notifies the orchestrator to update routing rules to steer traffic through the new replica.
For simultaneous failures, the orchestrator waits until all new replicas confirm that they have finished their state recovery procedures before updating routing rules.

\section{Implementation}
\label{section:implementation}
FTC builds on ONOS SDN controller~\cite{onos} and Click~\cite{click}.
The forwarder and buffer are implemented as Click elements.

A replica consists of {\em control} and {\em data plane} modules.
The control module is a daemon that communicates with the orchestrator and the control modules in other replicas.
In failover, the control module spawn a thread to fetch state per each replication group.
Using a reliable TCP connection, the thread sends a fetch request to the appropriate member in the replication group and waits to receive state.

The data plane module processes piggyback messages, sends and receives packets to and from a middlebox, constructs piggyback messages, and forwards packets to a next element in the chain (the data-plane module of the next replica or the buffer).

FTC appends the piggyback logs to the end of a packet, and inserts an IP option to notify our runtime that a packet has a piggyback message.
As a piggyback message is appended at the end of a packet, its process and construction can be performed in-place, and there is no need to actually strip and reattach it.
Before sending a packet to the middlebox, the relevant header fields (e.g., the total length in IP header) is updated to not accounting for the piggyback message. Before forwarding the packet to next replica, the header is updated back to reconsider the piggyback message.
For middleboxes that may extend the packet, the data plane module operates on the copy of a piggyback message.

\section{Evaluation}
\label{section:evaluation}
We describe our setup and methodology in \sect{setup}.
We benchmark the impact of state size on the performance of FTC in \sect{micro-benchmark}.
We measure the performance of FTC
for middleboxes in \sect{performance-middleboxes}
and for chains in \sect{ftsfc-in-normal-operation}.
Finally, we evaluate the failure recovery of FTC in \sect{ftsfc-in-failure-recovery}.

\subsection{Experimental Setup and Methodology}
\label{section:setup}

\begin{table}[t]
\setlength\tabcolsep{2pt}
    \centering
    \scriptsize
    \begin{tabular}{r l l r l}
        \cmidrule(lr){1-3}
        \cmidrule(lr){4-5}
        Middlebox
        &
        State reads
        &
        State writes
        &
        Chain
        &
        Middleboxes in chain
        \\
        \cmidrule(lr){1-3}
        \cmidrule(lr){4-5}
        \texttt{MazuNAT}
        &
        Per packet
        &
        Per flow
        &
        \texttt{Ch}-$n$
        &
        $\texttt{Monitor}_1 \rightarrow \dots \rightarrow \texttt{Monitor}_n$
        \\
        \texttt{SimpleNAT}
        &
        Per packet
        &
        Per flow
        &
        \texttt{Ch-Gen}
        &
        $\texttt{Gen}_1\rightarrow\texttt{Gen}_2$
        \\
        \texttt{Monitor}
        &
        Per packet
        &
        Per packet
        &
        \texttt{Ch-Rec}
        &
        \texttt{Firewall} $\rightarrow$ \texttt{Monitor} $\rightarrow$ \texttt{SimpleNAT}
        \\
        \texttt{Gen}
        &
        No
        &
        Per packet
        &
        &
        \\
        \texttt{Firewall}
        &
        N/A
        &
        N/A
        &
        &
    \end{tabular}
    \caption{Experimental middleboxes and chains}
    \label{table:implemented-stuff}
\end{table}

We compare FTC with \texttt{NF}, a non fault-tolerant baseline system, and \texttt{FTMB}, our implementation of \cite{rollback}.
Our \texttt{FTMB} implementation is a performance upper bound of the original work that performs the logging operations described in \cite{rollback} but does not take snapshots.
Following the original prototype, \texttt{FTMB} dedicates a server in which a middlebox \emph{master} (M) runs, and another server where the fault tolerant components \emph{input logger} (IL) and \emph{output logger} (OL) execute. Packets go through IL, M, then OL.
M tracks accesses to shared state using packet access logs (PALs) and transmits them to OL.
In the original prototype, no data packet is released until all corresponding dropped PALs are retransmitted. Our prototype assumes that PALs are delivered on the first attempt, and packets are released immediately afterwards. Further, OL maintains only the last PAL.

We used two environments.
The first is a local cluster of 12 servers. Each server has an 8-core Intel Xeon CPU D-1540 clocked at 2.0~Ghz, 64~GiB of memory, and two NICs, a 40~Gbps Mellanox ConnectX-3 and a 10~Gbps Intel Ethernet Connection X557.
The servers run Ubuntu 14.04 with kernel 4.4 and are connected to 10 and 40~Gbps top-of-rack switches.
We use MoonGen~\cite{MoonGen} and pktgen~\cite{pktgen} to generate traffic and measure latency and throughput, respectively.
Traffic from the generator server, passed in the 40~Gbps links, is sent through middleboxes and back to the generator. 
FTC uses a 10~Gbps link to disseminate state changes from buffer to forwarder.

The second environment is a distributed Cloud comprised of several core and edge data-centers deployed across Canada.
We use virtual machines with 4 virtual processor cores and 8~GiB memory running Ubuntu 14.04 with Kernel 4.4.
We use the published ONOS docker container~\cite{onos-docker} to control a virtual network of OVS switches~\cite{ovs} connecting these virtual machines.
We follow the \emph{multiple interleaved trials methodology} \cite{cloud-repeatable} to reduce the variability that come from performing experiments on a shared infrastructure.

\begin{figure*}[!ht]
	\centering
	\begin{minipage}[t]{0.33\linewidth}
	    \centering
        \includegraphics[width=\textwidth]{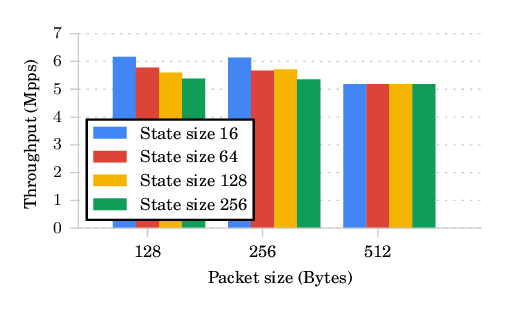}
        \caption{Throughput vs. state size}
        \label{fig:throughput-size}
	\end{minipage}
	\begin{minipage}[t]{0.33\linewidth}
		\centering
	    \includegraphics[width=\textwidth]{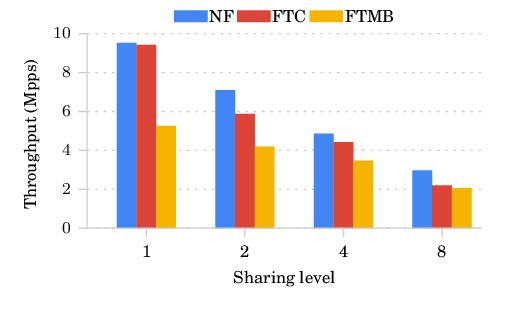}
        \caption{Throughput of \texttt{Monitor}}
        \label{fig:throughput-monitor}
	\end{minipage}
	\begin{minipage}[t]{0.33\linewidth}
		\centering
		\includegraphics[width=\textwidth]{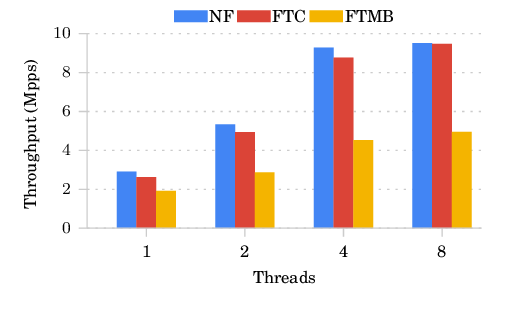}
        \caption{Throughput of \texttt{MazuNAT}}
        \label{fig:throughput-nat}
	\end{minipage}
\end{figure*}

We use the middleboxes and chains shown in \tabl{implemented-stuff}. The middleboxes are implemented in Click~\cite{click}.
\texttt{MazuNAT} is an implementation of the core parts of a commercial NAT  \cite{mazunat-click}, and \texttt{SimpleNAT}
provides basic NAT functionalities.
They represent read-heavy middleboxes with a moderate write load on the shared state.
\texttt{Monitor} is a read/write heavy middlebox that counts the number of packets in a flow or across flows.
It takes a \emph{sharing level} parameter that specifies the number of threads sharing the same state variable.
For example, no state is shared for the sharing level 1, and all 8 threads share the same state variable for sharing level 8.
\texttt{Gen} represents a write-heavy middlebox that takes a state size parameter, which allows us to test the impact of a middlebox's state size on performance.
\texttt{Firewall} is stateless.
Our experiments also test three chains comprised of these middleboxes, namely $\texttt{Ch-}n$, \texttt{Ch-Gen}, and \texttt{Ch-Rec}.

For experiments in the first environment, we report latency and throughput. 
For a latency data-point, we report the average of
hundreds of samples taken in a 10 second interval.
For a throughput data-point, we report the average of maximum throughput values measured every second in a 10-second interval.
Unless shown, we do not report confidence intervals
as they are negligible.
Unless told otherwise, the packet size in our experiments is 256~B, and $f=1$.

\subsection{Micro-benchmark}
\label{section:micro-benchmark}
We use a micro-benchmark to determine the impact of a state size on the performance of FTC.
We measured the latency overhead for the middlebox \texttt{Gen} and the chain \texttt{Ch-Gen}. We observed that under 2~Mpps for 512~B packets, varying the size of the generated state from 32--256~B has a negligible impact on latency for both \texttt{Gen} and \texttt{Ch-Gen} (the difference is less than 2~$\mu$s). Thus, we focus on the throughput overhead.

\begin{figure*}[!ht]
    \begin{subfigure}[l]{0.33\textwidth}
        \centering
        \includegraphics[width=\textwidth]{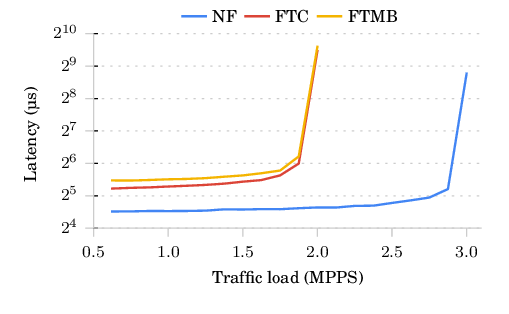}
        \caption{\texttt{Monitor} - Sharing level 8}
        \label{fig:latency-monitor}
    \end{subfigure}
    \begin{subfigure}[l]{0.33\textwidth}
        \centering
        \includegraphics[width=\textwidth]{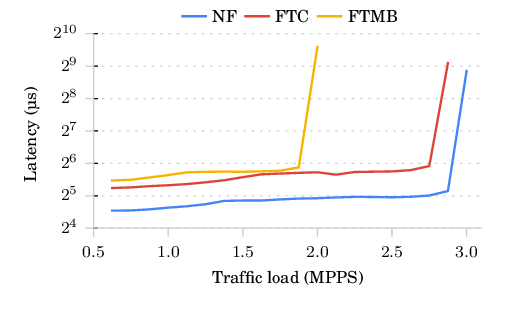}
        \caption{\texttt{MazuNAT} - 1 thread}
        \label{fig:latency-mazunat-1}
    \end{subfigure}
    \begin{subfigure}[l]{0.33\textwidth}
        \centering
        \includegraphics[width=\textwidth]{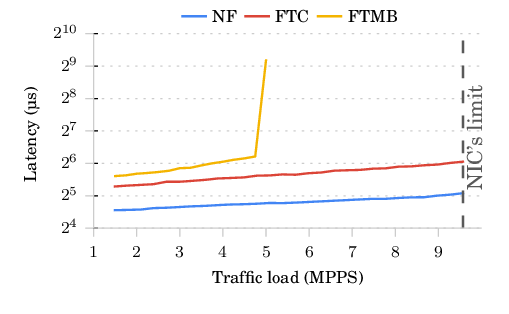}
        \caption{\texttt{MazuNAT} - 8 threads}
        \label{fig:latency-mazunat-8}
    \end{subfigure}
    \caption{Latency of middleboxes
    }
    \label{fig:latency-mbs}
\end{figure*}

\paragraph{Throughput:}
\fig{throughput-size} shows the impact of state size generated by \texttt{Gen} on throughput. \texttt{Gen} runs a single thread.
We vary the state size and measure \texttt{Gen}'s throughput for different packet sizes.
As expected, the size of piggyback messages impacts the throughput only if it is proportionally large compared to packet sizes. For 128~B packets, throughput drops by only 9\% when \texttt{Gen} generates states that are 128~B in size or less. 
The throughput drops by less than 1\% with 512~B packets and state up to 256~B in size.

We expect popular middleboxes to generate state much smaller than some of our tested values. For instance, a load balancer and a NAT generate a record per traffic flow \cite{modeling-middleboxes,rfc3234,rfc3022} that is roughly 32~B in size ($2 \times 12$~B for the IPv4 headers in both directions and 8~B for the flow identifier).

\subsection{Fault-Tolerant Middleboxes}
\label{section:performance-middleboxes}
\textbf{Throughput:}
Figures \ref{fig:throughput-monitor} and \ref{fig:throughput-nat} show the maximum throughput of two middleboxes.
In \fig{throughput-monitor}, we configure \texttt{Monitor} to run with eight threads and measure its throughput with different sharing levels. As the sharing level for \texttt{Monitor} increases, the throughput of all systems, including \texttt{NF}, drops due to the higher contention in reading and writing the shared state. For sharing levels of 8 and 2, FTC achieves a throughput that is $1.2\times$ and $1.4\times$ that of \texttt{FTMB}'s and incurs an overhead of 9\% and 26\% compared to \texttt{NF}.
These overheads are expected since \texttt{Monitor} is a write-heavy middlebox, and the shared state is modified non-deterministically per packet.
For sharing level 1, \texttt{NF} and FTC reach the NIC's packet processing capacity\footnote{Although the 40~GbE link is not saturated, our investigation showed that the bottleneck is the NIC's packet processing power. We measured that the Mellanox ConnectX-3 MT 27500, at the receiving side and working under the DPDK driver, at most can process 9.6--10.6~Mpps for varied packet sizes. Though we have not found any official document by Mellanox describing this limitation, similar behavior (at higher rates) has been reported for Intel NICs (see Sections 5.4 and 7.5 in \cite{MoonGen} and Section 4.6 in \cite{packet-shaker}).}.
\texttt{FTMB} does not scale for sharing level 1, since for every data packet, a PAL is transmitted in a separate message, which limits \texttt{FTMB}'s throughput to 5.26~Mpps.

\fig{throughput-nat} shows our evaluation for \texttt{MazuNAT}'s throughput while varying the number of threads.
FTC's throughput is 1.37--1.94$\times$ that of \texttt{FTMB}'s for 1 to 4 threads. Once a traffic flow is recorded in the NAT flow table, processing next packets of this flow only requires reading the shared record (until the connection terminates or times out).
The higher throughput compared for \texttt{MazuNAT} is because FTC does not replicate the reads, while \texttt{FTMB} logs them to provide fault tolerance~\cite{rollback}.
We observe that FTC incurs 1--10\% throughput overhead compared to \texttt{NF}. Part of this overhead is because FTC has to pay the cost of adding space to packets for possible state writes, even when state writes are not performed.

The pattern of state reads and writes impacts FTC's throughput. Under moderate write workloads, FTC incurs 1--10\% throughput overhead, while under write-heavy workloads, FTC's overhead remains less than 26\%.

\begin{figure*}[!ht]
	\centering
	\begin{minipage}[!ht]{0.33\linewidth}
		\centering
	    \includegraphics[width=\textwidth]{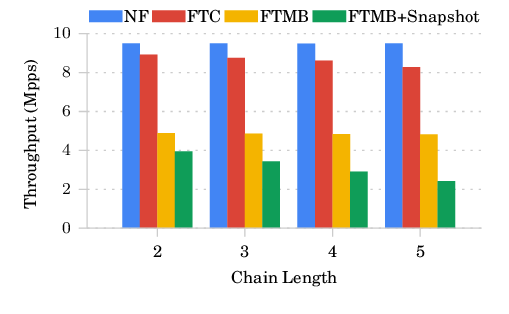}
		\caption{Tput vs. chain length}
        \label{fig:throughput-len}
	\end{minipage}
	\begin{minipage}[!ht]{0.33\linewidth}
		\centering
		\includegraphics[width=\textwidth]{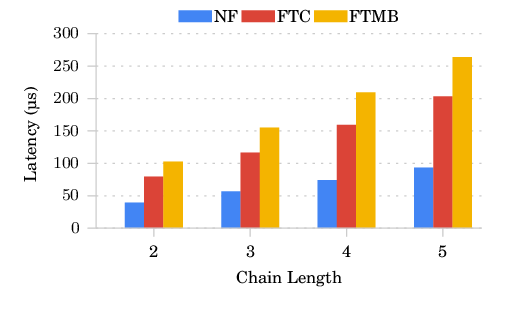}
		 \caption{Latency vs. chain length}
        \label{fig:latency-len}
	\end{minipage}
	\begin{minipage}[!ht]{0.33\linewidth}
		\centering
		\includegraphics[width=\textwidth]{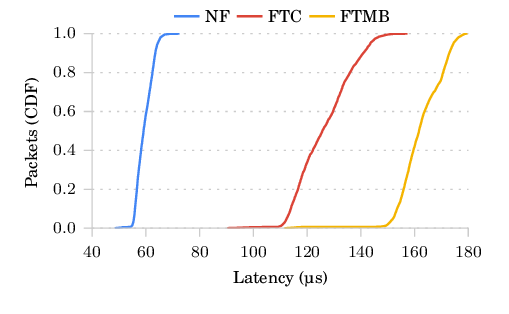}
		 \caption{\texttt{Ch-3} per packet latency}
        \label{fig:latency-len-cdf}
	\end{minipage}
\end{figure*}

\paragraph{Latency:}
\fig{latency-mbs} shows the latency of \texttt{Monitor} (8 threads with sharing level 8) and \texttt{MazuNAT} (two configurations, 1 thread and 8 threads) under different traffic loads.
For the both middleboxes, the latency remains under $0.7$~ms for all systems as the traffic load increases, until the systems reach their respective saturation points.
Past these points, packets start to be queued, and per-packet latency rapidly spikes.

As shown in \fig{latency-monitor}, under sustainable loads, FTC and \texttt{FTMB} respectively introduce overhead within 14--25~$\mu$s and 22--31~$\mu$s to the per packet latency, out of which 6--7~$\mu$s is due to the extra one way network latency to forward the packet and state to the replica.
For this write heavy middlebox, FTC adds a smaller latency overhead compared to \texttt{FTMB}.

\fig{latency-mazunat-1} shows that, when running \texttt{MazuNAT} with one thread, FTC can sustain nearly the same traffic load as \texttt{NF}, and FTC and \texttt{FTMB} have similar latencies. 
For eight threads shown in \fig{latency-mazunat-8}, both FTC and \texttt{NF} reach the packet processing capacity of the NIC. The latency of FTC is largely independent of the number of threads, while \texttt{FTMB} experiences a latency increase of 24--43~$\mu$s when going from one to eight threads.

\subsection{Fault Tolerant Chains}
\label{section:ftsfc-in-normal-operation}
In this section, we report the performance of FTC for a chain of middleboxes during normal operation.
For a \texttt{NF} chain, each middlebox is deployed in a separate physical server. We does not need more servers, while we dedicate twice the number of servers to \texttt{FTMB}: A server for each middlebox (Master in \texttt{FTMB}) and a server for its replica (IL and OL in \texttt{FTMB}).

\paragraph{Chain length impact on throughput:}
\fig{throughput-len} shows the maximum traffic throughput passing in four chains (\texttt{Ch-2} to \texttt{Ch-5} listed in \tabl{implemented-stuff}). \texttt{Monitor}s in these chains run eight threads with sharing level 1.
We also report for \texttt{FTMB+Snapshot} that is \texttt{FTMB} with snapshot simulation. To simulate the overhead of periodic snapshots, we add an artificial delay (6~ms) periodically (every 50~ms). We get these values from \cite{rollback}.

As shown in \fig{throughput-len}, FTC's throughput is within 8.28--8.92~Mpps and 4.83--4.80~Mpps for \texttt{FTMB}. FTC imposes a 6--13\% throughput overhead compared to \texttt{NF}.
The throughput drop from increasing the chain length for FTC is within 2--7\%, while that of \texttt{FTMB+Snapshot} is 13--39\% (its throughput drops from 3.94 to 2.42~Mpps).

This shows that throughput of FTC is largely independent of the chain length, while, for \texttt{FTMB+Snapshot}, periodic snapshots taken at all middleboxes significantly reduce the throughput. No packet is processed during a snapshot. Packet queues get full at early snapshots and remain full afterwards because the incoming traffic load is at the same rate. More snapshots are taken in a longer chain. Non-overlapping (in time) snapshots cause shorter service time at each period and consequently higher throughput drops.
An optimum scheduling to synchronize snapshots across the chain can reduce this overhead; however, this is not trivial \cite{unreliable-failure-detectors}.

\paragraph{Chain length impact on latency:}
We use the same experimental settings as the previous experiment, except we run single threaded \texttt{Monitor}s due to a limitation of the traffic generator. The latter is not able to measure the latency of the chain beyond size 2 composed of multithreaded middleboxes. We resort to use single threaded \texttt{Monitor}s under the load of 2~Mpps, a sustainable load by all systems.

As shown in \fig{latency-len}, FTC's overhead compared to \texttt{NF} is within 39--104~$\mu$s for \texttt{Ch-2} to \texttt{Ch-5}, translating to roughly 20$~\mu$s latency per middlebox. The overhead of \texttt{FTMB} is within 64--171~$\mu$s, approximately 35~$\mu$s latency overhead per middlebox in the chain.
As shown in \fig{latency-len-cdf}, the tail latency of individual packets passing through \texttt{Ch-3} is only moderately higher than the minimum latency.
FTC incurs $16.5$--$20.6$~$\mu$s per middlebox latency which is respectively three and two orders of magnitudes less than Pico's and REINFORCE's, and is around $2/3$ of \texttt{FTMB}'s.

In-chain replication eliminates the communication overhead with remote replicas.
Doing so also does not cause latency spikes unlike snapshot-based systems.
In FTC, packets experience constant latency, while the original FTMB reports up to 6~ms latency spikes at periodic checkpoints (e.g., at every 50~ms intervals) \cite{rollback}.

\begin{figure}[t]
	\centering
	\begin{minipage}[!ht]{0.49\linewidth}
		\centering
    	\includegraphics[width=\textwidth]{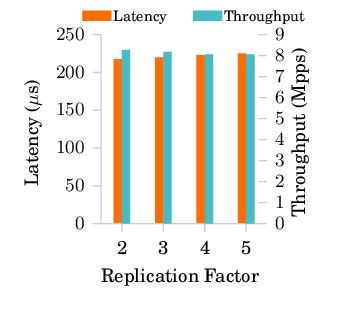}
        \caption{Repl. factor}
        \label{fig:perf-rep-factor}
	\end{minipage}
	\begin{minipage}[!ht]{0.49\linewidth}
        \centering
        \includegraphics[width=\textwidth]{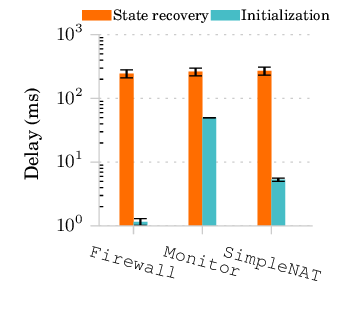}
        \caption{Recovery time}
        \label{fig:wan-recovery-time}
	\end{minipage}
\end{figure}

\paragraph{Replication factor impact on performance:}
For replication factors of 2--5 (i.e., tolerating 1 to 5 failures), \fig{perf-rep-factor} shows FTC's performance for \texttt{Ch-5} in two settings where \texttt{Monitor}s run with 1 or 8 threads. We report the throughput of 8 threaded \texttt{Monitor}, while only report the latency of 1 threaded \texttt{Monitor} due to a limitation of our test harness.

To tolerate $2.5\times$ failures, FTC incurs only 3\% throughput overhead as its throughput decreases to $8.06$~Mpps. The latency overhead is also insignificant as latency only increases by 8~$\mu$s.
By exploiting the chain structure, FTC can tolerate a higher number of failures without sacrificing  performance.
However, the replication factor cannot be arbitrarily large as encompassing the resulting large piggyback messages inside packets becomes impractical.

\subsection{FTC in Failure Recovery}
\label{section:ftsfc-in-failure-recovery}
Recall from \sect{implementation}, failure recovery is performed in three steps that incure initialization, state recovery, and rerouting delays.
To evaluate FTC during recovery, we measure the recovery time of \texttt{Ch-Rec} (see \tabl{implemented-stuff}). Each middlebox is placed in a different region of our Cloud testbed. As the orch detects a failure, a new replica is placed in the same region as the failed middlebox. The head of \texttt{Firewall} is deployed in the same region as the orch, while the heads of \texttt{SimpleNAT} and \texttt{Monitor} are respectively deployed in a neighboring region and a remote region compared to the orch's region. Since the orch is also a SDN controller, we observe negligible values for the rerouting delay, thus we focus on the state recovery delay and initialization delay.

\paragraph{Recovery time:}
As shown in \fig{wan-recovery-time}, the initialization delays are 1.2, 49.8, and 5.3~ms for \texttt{Firewall}, \texttt{Monitor}, and \texttt{SimpleNAT}, respectively.
The longer the distance between the orch and the new replica, the higher the initialization delay.
The state recovery delays are in the range of $114.38\pm9.38$~ms to $270.79\pm50.47$~ms\footnote{The large confidence intervals reported are due to latency variability in the wide area network connecting different regions.}.
In a local area network, FTMB paper~\cite{rollback} reports comparable recovery time of $\sim$100~ms to 250~ms for \texttt{SimpleNAT}.
The WAN latency between two remote regions becomes the dominant delay during failover, because a new replica fetches state from a remote region upon any failure.
Using ping, we measured the network delay between all pairs of remote regions, and the observed round-trip times confirmed our results.

FTC replicates the values of state variables, and its state recovery delay is bounded by the state size of a middlebox.
The replication factor also has a negligible impact on the recovery time of FTC, since a new instantiated replica fetches state in parallel from other replicas.
 
\section{Related Work}
\label{section:related-work}
We already discussed NFV related work in \sect{fault-tolerance-for-middleboxes}. Next, we position FTC related to three lines of work.

\paragraph{Fault tolerant storage:}
Prior to FTC, the distributed system literature used chain and ring structures to provide fault tolerance.
However,
their focus
is on ordering read/write messages at the process level (compared to, middlebox threads racing to access shared state in our case), at lower non-determinism rates (compared to, per-packet frequency), and at lower output rates (compared to, several Mpps releases).

A class of systems adapt the chain replication protocol~\cite{chain-replication}
for key-value storages. In HyperDex \cite{hyperdex} and Hibari \cite{hibari}, servers shape multiple logical chains replicating different key ranges.
NetChain \cite{netchain} replicates in the network on a chain of programmable switches.
FAWN \cite{fawn}, Flex-KV \cite{flex-kv}, and parameter server \cite{parameter-server} leverage consistent hashing to form a replication ring of servers.
Unlike these systems, FTC takes advantage of the natural structure of service function chains, uses transactional packet processing, and piggybacks state updates on packets.

\paragraph{Primary backup replication:}
In \emph{active} replication \cite{state-machine-replication}, all replicas process requests. 
This scheme requires determinism in middlebox operations, while middleboxes are non-deterministic \cite{routerbricks, packet-shaker}.
In passive replication \cite{primary-backup}, only a \emph{primary} server processes requests and sends state updates to other replicas.
This scheme makes no assumption about determinism.
Generic virtual machine high availability solutions~\cite{scales2010design,remus,colo}
pause a virtual machine per each checkpoint.
These solutions are not effective for chains, since the chain operations pauses during long checkpoints.

\paragraph{Consensus protocols:}
Classical consensus protocols, such as Paxos \cite{lamport2001paxos} and Raft \cite{raft} are known to be slow and cause unacceptable low performance if used for middleboxes.

\section{Conclusion}
\label{section:conclusion-and-future-works}
Existing fault tolerant middlebox frameworks can introduce high performance penalties when they are used for a service function chain.
This paper presented FTC, a system that takes advantage of the structure of a chain to provide efficient fault tolerance.
Our evaluation demonstrates that FTC can provide high degrees of fault tolerance with low overhead in terms of latency and throughput of a chain.

\bibliographystyle{plain}
\bibliography{library}

\end{document}